\documentclass[pre,aps,showpacs]{revtex4}

\newcommand{\beq}{\begin{equation}}
\newcommand{\eeq}{\end{equation}}
\newcommand{\bqn}{\begin{eqnarray}}
\newcommand{\eqn}{\end{eqnarray}}
\newcommand{\bqns}{\begin{eqnarray*}}
\newcommand{\eqns}{\end{eqnarray*}}
\newcommand{\bary}{\begin{array}}
\newcommand{\eary}{\end{array}}
\newcommand{\non}{\nonumber}

\usepackage{graphicx}%
\usepackage{amsmath}

\begin{document}
\title{\bf Analysis on the imaging properties of a left-handed material slab}
\author{Pi-Gang Luan, Hung-Da Chien, Chii-Chang Chen}
\affiliation{Institute of Optical Sciences, National Central
University, Chung-Li 32054, Taiwan, Republic of China }
\author{Chi-Shung Tang}
\affiliation{Physics Division, National Center for Theoretical Sciences,
P.O. Box 2-131, Hsinchu 30013, Taiwan, Republic of China}
\date{\today}

\begin{abstract}
We investigate in this paper the imaging properties of an absorptive
left-handed material (LHM) slab. For a line source, a geometric explanation to
the reason of the thickness limitation on an ideal lossless slab is given.
For a lossy slab, the imaging properties are determined by the wavelength, the slab thickness,
the distance from the source to the nearer boundary of the slab, and the absorption effect.
Varying the ratios between these quantities, the image width can be changed from wavelength to
subwavelength scale. In the former situation, the energy density is mainly concentrated at the
two image spots. In the later case, though image of subwavelength width appears on the focal
plane, however, most energy is located at about the two boundaries of the slab. The relations
between the subwavelength imaging and uncertainty principle are also discussed.

\end{abstract}

\maketitle

\section{Introduction}

Negative refraction of electromagnetic waves by a left-handed material (LHM),
first proposed in 1960s by Veselago \cite{1}, has attracted strong research
interests \cite{2,3,4,5,6} and generated heated debate\cite{7,8,9,10,11}.
Among all the phenomena that could happen in a LHM, the most
fascinating one may be the possibility of ``superlensing effect" proposed by
Pendry \cite{2}; that is, a slab made of uniform and isotropic LHM \cite{1} with both
the permittivity $\varepsilon=-1$  and the permeability $\mu=-1$ acquires a negative
refractive index $n=-1$, which makes this slab a perfect lens.  It can capture
both the propagating and the evanescent waves emitted from a point source placed in front
of the slab and refocuses them into two point images,
one inside and the other behind the slab.

Recently, this superlensing effect was questioned by a number of authors \cite{7,8,9,10,11}.
In Ref. \cite{9}, the authors augured that negative refraction of energy flow implies
the violation
of causality principle, and a little amount of absorption will largely
deform the waves. In Ref. \cite{10}, the authors showed that although there is
amplification of evanescent waves in an ideal lossless left-handed medium, however,
to avoid the divergence of the field energy inside the lens, it must be limited
to a thickness smaller than the distance between the line source and the nearer
boundary of the slab, thus perfect imaging is impossble. In addition, a little absorption
may destroy the negative refraction effect completely.
It was then found
that to make a left-handed material physically realizable, the medium must be
dispersive or absorptive. In Ref. \cite{11}, the recovery rate for a lossy slab was studied,
and the author showed that the image
quality can be significantly affected by the absorption effect. In Ref. \cite{12,13},
the authors showed that the energy flow indeed goes to the ``negative way" when passing
through the surface of an absorptive and dispersive LHM. In \cite{14},
a slab lens of photonic crystal was considered, and the simulation showed that negative
refraction of energy flow does not contradict the causality principle. Further in \cite{15,16},
the concept of ``constant frequency curves" introduced in \cite{6} were used to study the
refraction behavior of the waves in the the medium. Most interestingly, in \cite{16} an all-angle
negative refraction photonic crystal slab lens was designed to focus the light into a
subwavelength region.

Although the focusing effect of a LHM slab lens has already
been studied by a number of authors, however, in most previous studies researchers used
some Finite-Difference-Time-Domain (FDTD) method. The method is easy to implement but
the physical meanings of the simulation results are not easy to be extracted.
In some other studies the authors used frequency domain method, however, they usually
considered only one single Fourier component of the fields. To get
a definite result, one has to sum over these Fourier components.

In this paper we study the imaging problem using a spectrum
decomposition method. We first decompose the cylindrical wave emitted by a line source
into a series of plane waves of different transverse wave numbers. By considering the boundary
conditions at the source point and the two boundaries of the slab lens, we then can
determine the transmission and reflection coefficients for each plane wave. These
quantities are utilized to construct the field function in every space region.

Our method does not adopt complicated numerical skills, thus makes
us easier to get the physical insight. We also give a very simple geometrical explanation
to the reason of the thickness limitation for the ideal slab lens
(the $\varepsilon=\mu=n=-1$ case) \cite{10}. Finally, we found that
the imaging mechanism for a negative refraction lens system is subtler than
that of the conventional lens system.

\section{Model and Method}

We first describe the setup of the slab system. In this paper we consider only
the E-polarized wave,
which means that the wave propagation direction is parallel
to the XZ plane. The $x$ axis is
parallel to the two boundaries of the slab, and the boundary
near the source is the $z=0$ plane. A current line source
${\bf J}({\bf r})e^{-i\omega t}=
\hat{y}J_0\,\delta({\bf r}-{\bf r}_0)\,e^{-i\omega t}$
located at ${\bf r}_0=(x_0,z_0)=(0,z_0),\,z_0<0$, emits monochromatic waves of angular frequency
$\omega$, thus both the ${\bf E}$ and ${\bf H}$ fields get a time factor $e^{-i\omega t}$.
The ${\bf E}$ field wave radiated from it is ${\bf E}_{rad}({\bf r})e^{-i\omega t}
=\hat{y}A_0\,H^{(1)}_0(k|{\bf r}-{\bf r}_0|)\,e^{-i\omega t}$, which
satisfies
\beq
\left(\nabla^2+k^2\right){\bf E}_{rad}({\bf r})=-i\frac{4\pi\omega}{c^2} {\bf J}({\bf r}).
\eeq
Here $H^{(1)}_0(x)$
is the zeroth order Hankel function of the first kind, $J_0$ and $A_0=-\pi\omega J_0/c^2$
are two constants
propotional to each other, ${\bf r}$ is the
observation point, and $k=\omega/c$ and $c$ are the wave number of the cylindrical
wave and the speed of light in vacuum (outside of the slab), respectively.

To calculate the total ${\bf E}({\bf r})$ field, we first introduce the Green's function
satisfying
\beq \left(\nabla^2+k^2(z)\right)G({\bf r},{\bf r}')=-\delta^{(2)}({\bf r}-{\bf r}'),\eeq
then the ${\bf E}$ field is given by
\bqn {\bf E}({\bf r})&=&i\frac{4\pi\omega}{c^2}\int d^2 r'\, G({\bf r},{\bf r}')\,
{\bf J}({\bf r}')\non\\&=&i\frac{4\pi\omega}{c^2}J_0\,G({\bf r},{\bf
r}_0)\,\hat{y}.\eqn
Here $k^2(z)=k^2=\omega^2/c^2$ in the regions outside the slab, and
$k^2(z)=\varepsilon\mu\,\omega^2/c^2$ if $0\leq z \leq d$.
$\varepsilon$ and $\mu$ are the permmitivity and permeaility in the slab, respectively.

To proceed further, the waves have to be decomposed into various
Fourier components \cite{5}. Each component has a definite $k_x$. It is a plane wave with
either a real $k_z=\sqrt{\omega^2/c^2-k^2_x},$ if $|k_x|\leq\omega/c$, or an imaginary
$k_z$, if $|k_x|>\omega/c$. In the former case we have a propagating wave, and
in the later case the wave is evanescent.

Write $G({\bf r},{\bf r}_0)$ as
\beq G({\bf r},{\bf r}_0)=\frac{1}{2\pi}\int^{\infty}_{-\infty} dk_x\,e^{ik_x x}g(z,k_x),
\label{int}\eeq
then we have
\beq \left[ \frac{d^2}{dz^2}+k^2(z)-k^2_x\right]g(z,k_x)=-\delta(z-z_0),\eeq
which leads to the boundary condition for $g$ at $z=z_0$:
\beq g'(z,k_x)|_{z_0+}-g'(z,k_x)|_{z_0-}=-1.\eeq
The continunity conditions for the tangential components of
the ${\bf E}$ and ${\bf H}$ fields at the two boundaries of the slab lead to
\bqn g(z,k_x)|_{outside}&=& g(z,k_x)|_{inside},\\
g'(z,k_x)|_{outside}&=& \frac{1}{\mu}g'(z,k_x)|_{inside}.\eqn

Define
\beq \kappa_0=\sqrt{k^2-k^2_x},\;\;\;\;\;\kappa=\sqrt{k^2\varepsilon\mu-k^2_x},\eeq
the solution for $g$ is given by
\beq
g=\left\{\bary{lll}
\frac{e^{i\kappa_0|z-z_0|}+R\,e^{i\kappa_0(|z_0|-z)}}{-2i\kappa_0},\;\;z<0\\\\
\frac{e^{i\kappa_0|z_0|}\,T\,\left[\cos\kappa (z-d)
+i\frac{\mu\kappa_0}{\kappa}\sin\kappa(z-d)\right]}{-2i\kappa_0},\;\;\;0\leq z \leq d\\ \\
\frac{T\,e^{i\kappa_0(z-d+|z_0|)}}{-2i\kappa_0},\;\;\;\;z>d.\eary\right.
\eeq
Here
\beq
T=\frac{1}{\cos\kappa d -\frac{i}{2}\left(\frac{\kappa}{\mu \kappa_0}
+\frac{\mu\kappa_0}{\kappa}\right)\sin\kappa d}
\eeq
and
\beq R=\frac{\frac{i}{2}\left(\frac{\kappa}{\mu \kappa_0}
-\frac{\mu\kappa_0}{\kappa}\right)\sin\kappa d}
{\cos\kappa d -\frac{i}{2}\left(\frac{\kappa}{\mu \kappa_0}
+\frac{\mu\kappa_0}{\kappa}\right)\sin\kappa d}\eeq
are the transmission and reflection coefficients, calculated
from the transfer matrix method \cite{11,17}.

\section{An ideal Slab}

We now turn to the discussion of an ideal slab lens.
For an ideal slab we mean that we can find a frequency $\omega_0$ such that
for a dispersive medium slab lens medium with frequency dependent permmitivity
$\varepsilon(\omega)$ and permeability $\mu(\omega)$ and zero absorption effect
we have $\varepsilon(\omega_0)=\mu(\omega_0)=-1$.
Pendry pointed out in [2] that a slab lens of this kind is a perfect lens with $n=-1$.
It focuses the propagating waves and amplifies the evanescent waves,
thus can recover all the information carried by the wave emitted from
the line source. Although Pendry in his derivation showed that for a single
Fourier component the lens indeed amplifies the evanescent wave and thus the
amplitude of the wave can be completely recovered, however, he did not sum over these
Fourier components to get a result of the total field. In [10], the authors showed that
if the thickness $d$ of the lens is greater than $d_1=-z_0$, then the total field
will diverge inside of the lens. On the other hand, if $d<d_1$, there will be no image
at all. Thus perfect imaging is impossible.

Although the thickness limitation discussed in \cite{10} for an ideal
LHM slab lens is correct, however, it is hard to believe that there is some physical
principle that can restrict the slab thickness, if a thinner one can
be made. To resolve this puzzle, here we give a simple geometrical explanation to the
reason of this restriction (See Fig.1). Our explanation shows that the origin of the
restriction comes from the boundary conditions.

\begin{figure}[hbt]
\caption{\label{figure1}\small The field patterns as a function
of x and z for an ideal slab with $\varepsilon=\mu=-1$. The arrows indicate
the directions of energy flows. Inside the slab, energy flows in the direction opposite to
that of the wave vector ${\bf k}$. In (a) the slab has a thickness
$d$ shorter than the distance $d_1=|z_0|$ between the source point and the left surface
of the slab. The waves inside the slab region can be viewed of as radiated from a virtual source
behind the slab, whereas the waves in the region behind the slab can be viewed of as radiated
from a virtual source inside the slab. In (b) the slab has a thickness $d$ larger than $d_1=|z_0|$.
No stationary solution can exist in the empty (question marks) region.}
\end{figure}

Since the ideal slab
does not reflect light at all \cite{2}, thus the field inside and behind the
the slab are
\beq
E_{\rm inside}({\bf r},t)=A_0\,H^{(1)}_0(k|{\bf r}-{\bf r}_1|)\,e^{-i\omega t},\eeq
and
\beq
E_{\rm behind}({\bf r},t)=A_0\,H^{(1)}_0(k|{\bf r}-{\bf r}_2|)\,e^{-i\omega t},
\eeq
respectively. Here ${\bf r}_1=(0,-z_0)$ and ${\bf r}_2=(0,2d+z_0)=(0,2d-d_1)$ are the positions
of the two images predicted by the geometric optics.
Now, if $d<d_1$, then ${\bf r}_1$ and ${\bf r}_2$ are respectively located outside and inside
of the slab, respectively; that is, they are virtual images (virtual line sources).
In this case the
fields are finite everywhere except at the source point. However, if we increase
the slab thickness to $d>d_1$, then both images become real, and this contradicts
the boundary conditions. More specifically, a real image means a delta function term,
i.e., a line source. Since there is no any other line source
except the original one that located at ${\bf r}_0$, we conclude that the perfect
imaging is impossible.

Put it in another way. The time-averaged Poynting vector ${\bf S}$
must satisfy the divergenceless condition $\nabla\cdot {\bf S}=0$, thus there should be no
singular point satisfying $|{\bf S}|=\infty$ except the source point. Since in the slab the
wave vector
${\bf k}$ and ${\bf S}$ are antiparallel to each other, thus the waves propagating
in the $0<z<-z_0$ and $-z_0<z<d$ regions must be ``radiated from" and ``absorpted by"
the image inside the slab. This leads to the wave phase mismatch at $z=-z_0$ if $A_0\neq 0$.
From these consideration we conclude that the thickness limitation is a restriction
originating from the {\it boundary conditions} of this system, and it implies that
the stationary state (monochromatic waves) cannot satisfy these boundary conditions
simultaneously. In other words, {\it there is no stationary state}.

This result is consistent with the time domain results in Ref.\onlinecite{18},
where the source was treated as a driving force and the two surface plasmon modes were
two coupled oscillators. As one can see in Fig.3 of Ref.\onlinecite{18}, the
time evolution of the modulation amplitude $A(t)/A_{stat}(\tau\rightarrow\infty)$ oscillates
with a period $T_{osc}=4\pi/\Delta\omega_k$, where $\Delta\omega_k$ is the frequency difference
between the symmetric and antisymmetric surface plasmon modes. When the absorption of the slab
goes to zero, these two modes become degenerate, which leads to $T_{osc}\rightarrow\infty$
and $A_{stat}(\tau\rightarrow\infty)\rightarrow\infty$. This case
corresponds to the problem of driven oscillation without damping term. Therefore, the stationary
state will not appear, and the field energy inside the slab grows to a larger and larger value
without limitation.

\section{A Lossy Slab}

Now we turn to the discription of the numerical results for a lossy slab.
The permmitivity and permeability of the slab are chosen as
$\varepsilon=-1+i\delta_\varepsilon$ and $\mu=-1+i\delta_\mu$; both $\delta_\varepsilon$
and $\delta_\mu$ are small positive real numbers. With these parameters, the $g$ function
can be calculated. We then calculate the integral of Eq.(\ref{int}) numerically
as a sum. We first
let $k_x=k\tan\theta$, with $-f\pi/2<\theta <f\pi/2$. Here $\theta$ is
a reference angle, and $0<f<1$ gives the cutoff of $k_x$\cite{19}:
$(k_{x})_{max}=(\omega/c)\tan(f\pi/2)$. In this paper we choose $f=0.96$, which gives us
a $(k_{x})_{max}/k\approx 16$,
large enough and numerically implementable to give us meaningful results about subwavelength
imaging.
The range $(-f\pi/2,f\pi/2)$ is then being
discretized to $n_s=3000$ intervals, and the $dk_x$ is replaced by
$k\sec^2(\theta)d\theta$, with $d\theta=f\pi/n_s$.

\begin{figure}[hbt]
\caption{\label{figure1}\small (a1) The field strength pattern as a function
of $x$ and $z$. In this case $z_0=-1$, $d=2$, $\lambda=0.3$,
$\varepsilon=\mu=-1+0.001i$. The images have
widths of the wavelength scale. The two straight lines represent the boundaries
of the slab. (a2) The field strength at the focal plane as a function of $x$.
(a3) The field strength on the x=0 plane as a function of $z$. The three straight
lines represent the slab boundaries and the focal plane.
(b1) to (b3) are for the case of subwavelength images. In this case
$z_0=-1$, $d=2$, $\lambda=2$, and $\varepsilon=\mu=-1+0.000001i$.}
\end{figure}

Figure 2. shows two typical
cases for the imaging problem.
In case A (Fig.2(a1) to (a3)) the lens system creates two images,
one inside and one outside of the slab, and they have widths of the wavelength scale.
Here we have chosen $z_0=-1$, $d=2$, $\lambda=2\pi/k=0.3$, and
$\varepsilon=\mu=-1+0.001i$. We observe clearly that the largest field strength
locates at the two images. However, there is also
some surface resonance effect near the boundaries.
As we decrease the degree of the absorption, a stronger surface reresonance effect
is observed.
In case B (Fig.2(b1) to (b3)) we choose $z_0=-1$, $d=2$, $\lambda=2\pi/k=2$, and
$\varepsilon=\mu=-1+0.000001i$. In this case, the images become subwavelength scale.
It is also clear that the field strength is very large at the two boundaries of
the slab. This implies that surface-plasmon-polariton (SPP) plays important roles in this
case. It is interesting to note that, although on the focal plane the field strength
indeed has a peak along the $x$-direction, however,
the field strength does not have a local maximum around the image,
and in the $z$-direction the wave
strength decays from the second slab boundary. In this example the field strength
at the focal plane is about only $1\%$ of that at the boundaries. A closer observation
find that the field strength at the image point is the same order as that around
the source. This implies that if we turn on a line source, then the system has to spend
a long time (several hundreds of $2\pi/\omega$ or above) to build the energy
of the surface modes. Only after this transient process could the lens system focus the
light to a subwavelength space region.

The decaying profile of the field strength can be explained by the uncertainty principle.
According to this principle, we must have the relation
$\Delta x\,\Delta\,k_x\geq 1$, here the $\Delta x$ represents the width of
the image, and the $\Delta k_x$ represents the fluctuation of $k_x$. A
subwavelength image is mainly formed by summing over the Fourier components
of those $|k_x|\gg \omega/c$ terms. Since $k^2_z=\omega^2/c^2-k_x^2$, these components
must have imaginary $k_z$'s, and this leads to the decaying profile of the field strength.

An approximate image size can be obtained by analyzing the transmission
coefficient $T$. For $|k_x|\gg \omega/c$, and $\varepsilon=\mu=-1+i\delta$, $\delta<<1$,
we have
\beq T\approx
\frac{1}{e^{-|k_x|d}+\frac{\delta^2}{4}e^{|k_x|d}},\eeq
which is a hyperbolic secant function with a peak value
\beq T_{max}\approx\frac{1}{\delta}\eeq
at the transverse wave number
\beq\bar{k}_x=(1/d)\ln(2/\delta).\eeq Thus the image size is given by
\beq W=\frac{2\pi d}{\ln(2/\delta)}.\label{size}\eeq
A similar result has already been given by  Merlin in Ref.~\onlinecite{20}.

For the case B of Fig.2, we have $W\approx 0.43\lambda$, which is indeed a
subwavelength focusing. However, the actual size of the image is in fact a little
larger than that given by Eq.(\ref{size}). The reason is that the $g(z,k_x)$ function
in the $z>d$ region contains a factor $e^{\kappa_0(z-d+|z_0|)}/(-2i\kappa_0)$, and
thus in the integral (\ref{int}) the contributions from Fourier component with
$|k_x|<\bar{k}_x$ cannot be neglated. It seems that the near field excitations
(evascent surface waves) and small enough absorption play the most important roles in
the subwavelength imaging process.

\section{Conclusion}

In conclusion, we have studied the imaging properties of a negative-refraction slab lens,
using a spectrum decomposition method. We have also given a simple geometrical
explanation to the reason of the slab thickness limitation for an ideal negative
refraction lens. For a slab with appropriate amount absorption, we found that both
the wavelength size and subwavelength size images can be formed.

\section{ACKNOWLEDGMENT}

This work was suppoted by NSC, NCTS and NCU. Discussions
with Dr. D. H. Lin and Prof. Zhen Ye are also acknowledged.

\end{document}